# A REVIEW PAPER ON MICROPROCESSOR BASED CONTROLLER PROGRAMMING

Gurpreet Singh Sandhu,
*Senior Lecturer*
*GTBKIET, Malout ,Punjab*
gogimalout@gmail.com

Jaswinder Singh Dilawari
*Ph.D Research Scholar,*
*Pacific University, Udaipur, Rajasthan*
dilawari.jaswinder@gmail.com

*ABSTRACT: Designing of microprocessor based controllers requires specific hardware as well as software programming. Programming depends upon type of the software whether operating software or application software. Programming requires knowledge of system configuration and controller specific programming. Programs are always in digital form so microprocessor can control directly at digital level called Direct Digital Control (DDC).*

## 1. INTRODUCTION

In the early 1960 computer based controllers were used. They were having one main frame computer and all control action was dependent on it, moreover they were costly. But with the advent of microprocessor cost of controlling the plant decreased very less. In actual a microprocessor is a computer on a chip, and high-density memories reduced costs and package size dramatically and increased application flexibility. These controllers' measure signals from sensors, perform control routines in software programs, and take corrective action in the form of output signals to actuators. Since the programs are in digital form, the controllers perform what is known as direct digital control (DDC). Microprocessor can directly control the plant digitally. A direct digital control can be defined as the controller which updates the process as function of measured output variable and input provided. As the output world talks in analog form so for control digitally it has to be converted into digital form. For this A/D and D/A converters are used as shown in fig. 1

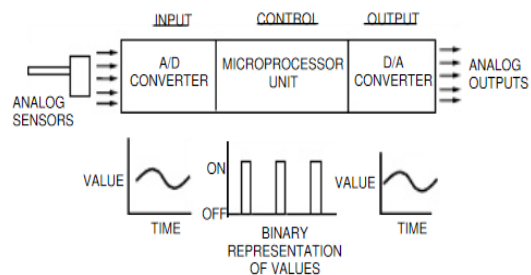

Figure1: A microprocessor based control system use A/D- D/A converter

A block diagram of microprocessor based digital control system along is shown in figure2 [1].

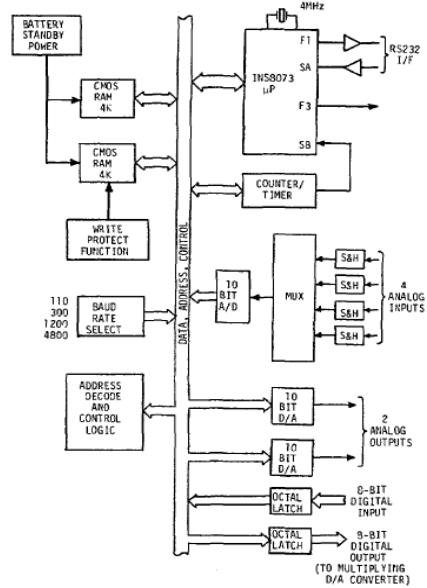

Figure2: Microprocessor based digital control system

Figure 2 shows the analog input and output through A/D and D/A converter.

## 2. CONTROLLER CONFIGURATION

The basic elements of a microprocessor-based (or micro-processor) controller (Fig.3) include:

—The microprocessor

—A program memory

—A working memory

—A clock or timing devices

—A means of getting data in and out of the system

In addition, a communications port is not only a desirable feature but a requirement for program tuning or interfacing with a central computer or building management system. Timing for microprocessor operation is provided by a battery-backed clock. The clock operates in the microsecond range controlling execution of program instructions. Program memory holds the basic instruction set for controller operation as well as for the application programs. Memory size and type vary depending on the application and whether the controller is considered a dedicated purpose or general purpose device. Dedicated purpose configurable controllers normally have standard programs and are furnished with read only memory (ROM) or programmable read only memory (PROM.). General purpose controllers often accommodate a variety of individual custom programs and are supplied with field-alterable memories such as electrically erasable, programmable, read only memory (EEPROM) or flash memory. Memories used to hold the program for a controller must be nonvolatile, that is, they retain the program data during power outages. A/D converters for DDC applications normally range from 8 to 12 bits depending on the application. An 8-bit A/D converter provides a resolution of one count in 256. A 12-bit A/D converter provides a resolution of one count in 4096. If the A/D converter is set up to provide a binary coded decimal (BCD) output, a 12-bit converter can provide values from 0 to 999, 0 to 99.9, or 0 to 9.99 depending on the decimal placement [3].

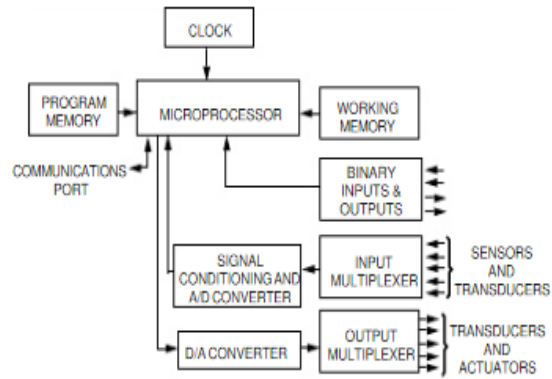

Figure3: Microprocessor Controller Configuration for automatic Control Applications

## 3. CONTROLLER SOFTWARE

Although use of microprocessor controller for any application depends upon the hardware but software determines the functionality. Controller software falls basically into two categories:

1. Operating software which controls the basic operation of the controller

2. Application software which addresses the unique control requirements of specific applications

### 3.1 Operating software

It is generally stored in volatile memory such as ROM, PROM. Operating software includes the operating system (OS) and routines for task scheduling, I/O scanning; priority interrupt processing, A/D and D/A conversion, and access and display of control program variables such as set points, temperature values, parameters, and data file information. Tasks are scheduled sequentially and interlaced with I/O scanning and other routine tasks in such a way as to make operation appear almost simultaneous[4]. If any higher priority task appears to operating software then current going task is ceased and data held in registers and accumulators are temporarily transferred to temporary registers. These interrupt requests are processed by priority interrupt register. When interrupt task is over then normal routine is resumed and data is transferred back from temporary registers to mainstream. The effect of these interrupts is transparent to the application that the controller is controlling

### 3.2 Application Software

Application software includes direct digital control, energy management, lighting control, and event initiated programs plus other alarm and monitoring software typically classified as building management functions. The system allows application programs to be used individually or in combination. For example, the same hardware and operating software can be used for a new or existing building control by using different programs to match the application. An existing building, for example, might require energy management software to be added to the existing control system. A new building, however, might require a combination of direct digital control and energy management software[5].

### 3.2.1 DIRECT DIGITAL CONTROL SOFTWARE

DDC software is used for specific control actions. These are set of standard DDC operators. Key elements in most direct digital control programs are the PID and the enhanced EPID and ANPID algorithms. While the P, PI, PID, EPID, and ANPID operators provide the basic control action, there are many other operators that enhance and extend the control program. Some other typical operators are shown in Table 1. These operators are computer statements that denote specific DDC operations to be performed in the controller [6]. Math, time/calendar, and other calculation routines (such as calculating an enthalpy value from inputs of temperature and humidity) are also required

**Table 1. Typical DDC Operators.**

| Operator | Description |
|---|---|
| Sequence | Allows several controller outputs to be sequenced, each one operating over a full output range. |
| Reversing | Allows the control output to be reversed to accommodate the action of a control valve or damper actuator. |
| Ratio | Translates an analog output on one scale to a proportional analog output on a different scale. |
| Analog controlled digital output | Allows a digital output to change when an analog input reaches an assigned value. Also has an assignable dead band feature. |
| Digital controlled analog output | Functionally similar to a signal switching relay. One state of the digital input selects one analog input as its analog output; the other state selects a second analog input as the analog output. |
| Analog controlled analog output | Similar to the digital controlled analog output except that the value and direction of the analog input selects one of the two analog signals for output. |
| Maximum input | Selects the highest of several analog input values as the analog output. |
| Minimum input | Selects the lowest of several analog input values as the analog output. |
| Delay | Provides a programmable time delay between execution of sections of program code. |
| Ramp | Converts fast analog step value changes into a gradual change. |

**4. CONTROLLER PROGRAMMING**

Controller programming makes the controller usable for a specific control action. Programming of microcomputer-based controllers can be subdivided into four discrete categories:

1. Configuration programming

2. System initialization programming

3. Data file programming

4. Custom control programming

Some controllers require all four levels of program entry while other controllers, used for standardized applications, require fewer levels.

Configuration programming matches the which hardware and software matches the control action required. It requires the selection of both hardware and software package to match the application requirement.

System initialization programming consists of entering appropriate startup values using a keypad or a keyboard. Star tup data parameters include set point, throttling range, gain, reset time, time of day, occupancy time, and night setback temperature[7]. These data are equivalent to the settings on a mechanical control system, but there are usually more items because of the added functionality of the digital control system.

Requirement of data file programming depends upon whether the system variables are fixed or variable. For example at zonal level programming where input sensors are fixed and programmer knows which relay will get output then the use of data file programming is irrelevant. But at the system level programming where controller controls wide variety of sensors and gives output to various relays, use of data file programming is must. For the controller to properly process input data, for example, it must know if the point type is analog or digital. If the point is analog, the controller must know the sensor type, the range, whether or not the input value is linear, whether or not alarm limits are assigned, what the high and low alarm limit values are if limits are assigned, and if there is a lockout point. See Table 2.If the point is digital, the controller must know its normal state (open or closed)[8], whether the given state is an alarm state or merely a status condition, and whether or not the condition triggers an event-initiated program.

Table 2. Typical Data File for Analog Input.

| Point Address | User Address |
|---|---|
| Point type | Regular or calculation |
| Sensor | Platinum (0 to 100F) |
| Physical terminal assigned | 16 |
| Use code | Cold deck dry bulb |
| Engineering unit | F |
| Decimal places for display | XXX.X |
| High limit | 70.0 |
| Low limit | 40.0 |
| Alarm lockout point | Point address |
| Point descriptor | Cold deck temperature |
| Alarm priority | Critical |

Custom control programming is the most involved programming category. Custom control programming requires a step-by-step procedure that closely resembles standard computer programming. A macro view of the basic tasks is shown in Figure 4.

## 5. CONCLUSION

Microprocessor based controllers although depends upon the hardware of controller but the main behavior is defined in software programming. Application software is used if a specific controlling action is needed. Before programming the controller values initial parameters is considered. Complexity of programming also depends upon the number of controllers to be controlled, input is analog or digital. If many inputs are coming to controller then a data file has to be maintained so that just by looking into that file constraints of programming can be identified.

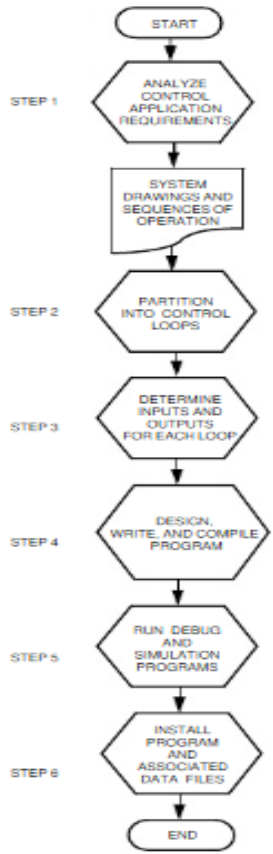

Figure4: Steps for custom level programming